\documentclass[aps,pra,amsmath,twocolumn]{revtex4}
\usepackage{graphicx}
\begin{document}
\draft
\title{Measuring the Density Matrix by Local Addressing}

\author{Z. Kis\thanks{Permanent address: Research Institute for
Solid State Physics and Optics, H-1525 Budapest, P.O. Box 49, Hungary
} and S. Stenholm}

\address{ Department of Physics, Royal Institute of Technology (KTH),
  Lindstedtsv\"agen 24, SE-10044 Stockholm, Sweden} 

\date{\today{}}

\begin{abstract}   
  We introduce a procedure to measure the density matrix of a material
  system. The density matrix is addressed locally in this scheme by
  applying a sequence of delayed light pulses. The procedure is based
  on the stimulated Raman adiabatic passage (STIRAP) technique. It is
  shown that a series of population measurements on the target state
  of the population transfer process yields unambiguous information
  about the populations and coherences of the addressed states, which
  therefore can be determined.
\end{abstract}

\maketitle

Active manipulation of the quantum state of different microscopic
systems increases the need for measurement schemes which verify the
reliability and efficiency of the engineering procedure. Quantum state
reconstruction methods have been proposed in several systems,
including light field and matter systems such as the vibrational state
of molecules, trapped atom motion, Bose-Einstein condensates, atomic
matter waves, electron motion etc., for recent reviews see
Refs.~\cite{ulf,vogel}.

In this work we introduce a procedure to measure the density matrix of
a material system. The density matrix is addressed locally in our
scheme. More precisely, the measurement process yields a part of the
total density matrix, namely the elements
\begin{equation}\label{rho}
  \left[\begin{array}{ccccc}
      & \vdots & & \vdots & \\
      \cdots & \varrho_{mm} & \cdots & \varrho_{mn} & \cdots \\
      & \vdots & & \vdots & \\
      \cdots & \varrho_{nm} & \cdots & \varrho_{nn} & \cdots \\
      & \vdots & & \vdots & 
      \end{array}\right],
\end{equation}
where the indices $m$ and $n$ refer to some steady states of the
material system. The procedure is based on population measurements, so
one needs an ensemble of identically prepared systems to obtain the
required density matrix elements. In practice this means that the
measurement is performed for example on an atomic or a molecular beam.
A population measurement inevitably entails a reduction of
the quantum state, therefore the state of the system is destroyed in
the process. The complete density matrix can be determined by repeated
application of the measurement procedure for different pairs of states
$\{m', n'\}$.  Recently, some new proposals have been published to
achieve a similar goal \cite{vit1,vit2}. We will compare our method
with those works at the end of this paper.

\begin{figure}
\includegraphics{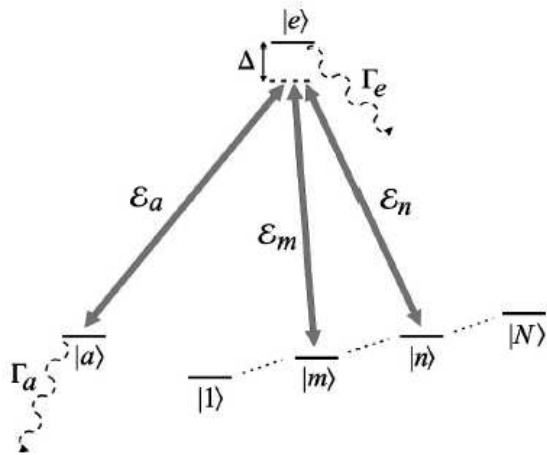}
\caption{Interaction scheme for the measurement of the density matrix.
  The states labeled by arabic numbers are occupied, the others are
  empty initially. The three laser pulses ${\cal E}_i(t)$ couple two of
  the populated states $|m\rangle$ and $|n\rangle$ with the auxiliary
  state $|a\rangle$ through the excited state $|e\rangle$. A common
  detuning $\Delta$ is allowed but the three photon resonance is
  required. The wavy arrows indicate decay from the excited and
  auxiliary states.}
\label{scheme}
\end{figure}

Our measurement procedure is based on the stimulated Raman adiabatic
passage (STIRAP) process \cite{review}.  Its implementation imposes
the following requirements on the system: the linkage displayed in
Fig.~\ref{scheme} must be realizable. We assume that the states
labeled by $\{|1\rangle, |2\rangle\, \ldots, |N\rangle\}$ are
populated, the others are empty. There are three light pulses $\{
{\cal E}_m(t), {\cal E}_n(t), {\cal E}_a(t)\}$ which couple the states
$\{|m\rangle, |n\rangle, |a\rangle\}$ via an excited state
$|e\rangle$. The other populated states must remain unaffected by
these light fields.  This could be achieved by exploiting selection
rules or selection based on resonance frequencies.  The frequencies of
the light fields may be detuned by $\Delta$ from the transition
frequencies $\omega_{ei}\!=\!(E_e-E_i)/\hbar$, where $E_i$ is the
energy of the state $|i\rangle\in\{|a\rangle, |1\rangle, \ldots,
|N\rangle\}$, but the three-photon resonance condition must be
fulfilled.

Initially, the states $|a\rangle$ and $|e\rangle$ are empty. The
Master equation which describes the measurement process reads
\begin{eqnarray}\label{mast}
  \frac{\partial \hat{\varrho}}{\partial t}\!=\!&-&\frac{i}{\hbar}
  [\hat H(t), \hat{\varrho}] -\frac{\varGamma_e}{2}
  (|e\rangle\langle e|\hat{\varrho} +
  \hat{\varrho} |e\rangle\langle e|) \nonumber \\
  &-& \frac{\varGamma_a}{2}
  ( |a\rangle\langle a|\hat{\varrho} +
  \hat{\varrho} |a\rangle\langle a|),
\end{eqnarray} 
where $\hat{\varrho}$ denotes the density operator of the system. The
constants $\varGamma_e$ and $\varGamma_a$ stand for the decay rates
from the states $|e\rangle$ and $|a\rangle$, respectively.
The Hamiltonian $\hat H(t)$ is given in the interaction picture and
the rotating-wave approximation (RWA) as
\begin{equation}\label{ham}
  \hat H(t)\!=\!\hbar\Delta|e\rangle\langle e|+\frac{\hbar}{2}
  \sum_{i=m, n, a}(\Omega_i(t)|i\rangle\langle e|+h.c.),
\end{equation} 
where the pulsed Rabi frequency $\Omega_i(t)$ derives from the field
${\cal E}_i(t)$. The Rabi frequencies $\Omega_m (t)$ and $\Omega_n
(t)$ are taken to have the same envelopes: 
\begin{equation}\label{omdef}
  \Omega_m(t)\!=\!\Omega_p(t)\cos\alpha, \qquad \Omega_n(t)\!=\!\Omega_p
  (t)\sin\alpha\,e^{i\beta},
\end{equation}
where $\alpha$ and $\beta$ are fixed angles which define the relative
amplitudes and phase of the two pulses, respectively.  The pulses $m,
n$ and $a$ are delayed with respect to each other; however, for an
efficient STIRAP process they must have a significant overlap.  In the
following $\Omega_p(t)$ and $\Omega_a(t)$ will be taken real.

Let us neglect the dissipative terms in the Master equation
(\ref{mast}). Then we have a purely unitary evolution. Now, our aim is
to determine the time evolution operator $\hat U(t)$ which governs the
time development of the density operator in the absence of
dissipation. It will help us understand the essence of the
measurement procedure, and we later return to the discussion of
dissipations.

The measurement procedure consists of performing a STIRAP process in
which a part of the population from the states $|m\rangle$ and
$|n\rangle$ is transferred to the state $|a\rangle$. The pulses $m,
n$, which act on the states $\{|m\rangle, |n\rangle\}$, play the role
of the pump pulses, whereas the pulse $a$ corresponds to the Stokes
pulse. It is assumed that they arrive in the counterintuitive time
order.  The pump pulses $m, n$ define a coupled state (or bright
state)
\begin{equation}\label{cc}
  |C\rangle\!=\!\cos\alpha\,|m\rangle + \sin\alpha\,e^{i\beta}|n\rangle,
\end{equation}
and an orthogonal decoupled state (or dark state)
\begin{equation}
  |D\rangle\!=\!-\sin\alpha\,|m\rangle + \cos\alpha\,e^{i\beta}|n\rangle,
\end{equation}
see Ref.~\cite{ari}. The Hamiltonian in Eq.~(\ref{ham}) can be
expressed in terms of the states $|C\rangle$ and $|D\rangle$ to obtain
\begin{equation}\label{hamnew}
  \hat H(t)\!=\!\hbar\Delta|e\rangle\langle e|+\frac{\hbar}{2}
  (\Omega_p(t)|C\rangle\langle e|+\Omega_a(t)|a\rangle\langle e|+h.c.).
\end{equation}
Obviously, the Hamiltonian does not act on the decoupled state
$|D\rangle$.  From this formulation it can clearly be seen that we have in
fact an ordinary STIRAP process defined on a three-state system.  The
dark state $|\psi_0(t)\rangle$ of the Hamiltonian in
Eq.~(\ref{hamnew}) is given by
\begin{equation}\label{dark}
  |\psi_0(t)\rangle\!=\cos\theta |C\rangle - \sin\theta |a\rangle,
\end{equation}
and the two bright states are
\begin{eqnarray}\label{bright}
  |\psi_{+}(t)\rangle\!&=&\sin\varphi\sin\theta |C\rangle + 
  \sin\varphi\cos\theta|a\rangle+ \cos\varphi |e\rangle 
  \nonumber \\
  \\
  |\psi_{-}(t)\rangle\!&=&\cos\varphi\sin\theta |C\rangle + 
  \cos\varphi\cos\theta|a\rangle- \sin\varphi |e\rangle,
  \nonumber
\end{eqnarray}
where the angles $\theta$ and $\varphi$ are defined as
\begin{equation}
  \tan\theta\!=\!\frac{\Omega_p(t)}{\Omega_a(t)},\qquad
  \tan2\varphi\!=\!\frac{\Omega(t)}{\Delta},
\end{equation}
with $\Omega(t)\!=\!\sqrt{\Omega_p^2(t)+\Omega_a^2(t)}$, see
Ref.~\cite{review}.  In the adiabatic limit, a simple calculation
shows that the unitary time evolution operator $\hat U(t)$ reads
\begin{eqnarray}
  \hat U(t)\!=&&|D\rangle\langle D|+|\psi_0(t)\rangle\langle C|
  \nonumber \\
  &&+\exp\left(-\frac{i}{\hbar}\int_{-\infty}^t\varepsilon_{+}(t')dt'\right)
  |\psi_{+}(t)\rangle \langle\psi_{+}(-\infty)|
  \nonumber \\
  &&+\exp\left(-\frac{i}{\hbar}\int_{-\infty}^t\varepsilon_{-}(t')dt'\right)
  |\psi_{-}(t)\rangle \langle\psi_{-}(-\infty)|,
\end{eqnarray} 
where $\varepsilon_{+}(t)\!=\!\frac{\hbar}{2}\Omega(t)\cot\varphi$ and
$\varepsilon_{-}(t)\!=\!-\frac{\hbar}{2}\Omega(t)\tan\varphi$ are the two
nonzero eigenenergies of the Hamiltonian Eq.~(\ref{hamnew}). The
physical interpretation of this result is quite simple: The decoupled
state $|D\rangle$ remains untouched throughout the transfer process.
The coupled state $|C\rangle$ follows adiabatically the dark state
$|\psi_0(t)\rangle$ of the Hamiltonian $\hat H(t)$
[Eq.~(\ref{hamnew})]. The bright states $|\psi_{\pm}(t)\rangle$ also
evolve adiabatically, however, they acquire a phase shift because they
belong to nonzero eigenenergies $\varepsilon_{\pm}(t)$.

In the adiabatic limit, the bright states $|\psi_{\pm}(t)\rangle$ are
not populated throughout the whole time, provided that initially
they were not populated. Thus, we have found that the final state of
the system, after the pulses have passed becomes
\begin{eqnarray}
  \lim_{t\rightarrow\infty}\hat U(t)\hat\varrho_i \hat U^{\dag}(t)
  &=&\!|D\rangle\langle D|\hat\varrho_i|D\rangle\langle D|
  +|a\rangle\langle C|\hat\varrho_i|C\rangle\langle a|
  \nonumber \\
  &-&|D\rangle\langle D|\hat\varrho_i|C\rangle\langle a|
  -|a\rangle\langle C|\hat\varrho_i|D\rangle\langle D|.
\end{eqnarray}

Let us suppose that we are able to measure the population $P_a$ on the
state $|a\rangle$. Then, the result can be expressed in terms of the
density matrix elements $(\hat\varrho_i)_{mm}$,
$(\hat\varrho_i)_{nn}$, $(\hat\varrho_i)_{mn}$ as
\begin{eqnarray}\label{Ps}
  P_a\!=\!\langle C|\hat\varrho_i|C\rangle \!&=&\!
  \cos^2\alpha\,(\hat\varrho_i)_{mm}+ 
  \sin^2\alpha\,  (\hat\varrho_i)_{nn} \nonumber \\
  &+&\sin 2\alpha\,\text{Re}\{(\hat\varrho_i)_{mn}\,e^{i\beta}\},
\end{eqnarray}
where we have made use of the definition Eq.~(\ref{cc}).  The outcome of
the measurement is phase sensitive: It depends on the relative phase
$\beta$ between the fields $m$ and $n$ (cf. Eq.~(\ref{omdef})).
Observe, that the coherence $(\hat\varrho_i)_{mn}$ also appears, so
one expects that by accomplishing an appropriate series of
measurements one can obtain not only the diagonal elements of the
density matrix, but the coherences as well. Indeed, the three required
density matrix elements can be determined in four steps: In the first
and second measurements, one of the pump field $n$ or $m$ is switched
off which corresponds to taking $\alpha\!=\!0$ or $\alpha\!=\!\pi/2$. In
this way, we get the diagonal density matrix elements
$(\hat\varrho_i)_{mm}$ and $(\hat\varrho_i)_{nn}$.  In the third
and fourth steps, both of the pump fields $m$ and $n$ are present so
that the value of $\alpha$ is chosen as $0\!<\!\alpha\!<\!\pi/2$. One
measurement is performed with the choice $\beta\!=\!0$ and an other
with $\beta\!=\!-\pi/2$. It is easy to show, that by combining the
results of the last two measurements with that of the first and
second, one can unambiguously determine the needed
$(\hat\varrho_i)_{mn}$.

There is one question left, which we have to answer: How can one
measure the population $P_a$? Here we return back to the analysis of
the dissipation mechanisms which we have postponed so far. Let us consider
again the Master equation (\ref{mast}). If the usual STIRAP conditions
are met during the measurement procedure, then the decay from the
excited state $|e\rangle$ has got negligible effect, since this state
is minimally populated \cite{glushko,vit3}. The situation is quite
different in the case of the decay from the target state $|a\rangle$:
This state is deeply involved in the transfer process. The dissipation
may influence significantly the population transfer to the target
state and as a result, it deteriorates the fidelity of the measurement.
However, it can also be utilized to monitor the population on the
target state $|a\rangle$.

It is plausible to assume, that the decay rate from the state
$|a\rangle$ to the states $|m\rangle$ and $|n\rangle$ is negligibly
small, since they have the same parity, because both of the transitions
$|m\rangle\leftrightarrow|e\rangle$ and
$|e\rangle\leftrightarrow|a\rangle$ are dipole allowed; the argument
is the same in the case of the state $|n\rangle$. The decay from the
state $|a\rangle$ to the state $|e\rangle$ is undesirable. It can be
avoided if the energy of the state $|a\rangle$ is lower than that of
the state $|e\rangle$ or if there are faster decay channels. In the
following we assume that the decay between these states is negligible.

Now we present a simplified treatment of the dissipation from the
target state $|a\rangle$ by assuming that the initial state of the
system can be given by a state vector $\pmb{B}\!=\![B_C, B_e, B_a]^T$
in the basis $\left\{|C\rangle, |e\rangle, |a\rangle\right\}$. Note
that in this definition we have included only those states of the system
which participate in the population transfer process. This means that
the norm of $\pmb{B}$ may be smaller than unity. The Master equation
(\ref{mast}) is replaced by a Schr\"odinger equation with a
non-Hermitian Hamiltonian. In the adiabatic basis Eqs.~(\ref{dark})
and (\ref{bright}), the Schr\"odinger equation reads
\begin{equation}\label{sch2}
  i\frac{d}{dt}\pmb{A}(t)=\pmb{H}'(t) \pmb{A}(t),
\end{equation}
where the effective non-Hermitian Hamiltonian $\pmb{H}'(t)$  is given by
\begin{widetext}
\begin{equation}\label{hamadi}
  \pmb{H}'(t)\!=\!\left[
    \begin{array}{ccccc}
      \frac{1}{2}\Omega\cot\varphi  
      -i\varGamma_a\cos^2\theta\sin^2\varphi & &
      i\left(\frac{1}{2}\varGamma_a\sin(2\theta) + 
        \dot{\theta}\right)\sin\varphi & &
      -i\frac{1}{2}\varGamma_a\cos^2\theta\sin(2\varphi)+i\dot{\varphi}
      \\ \\
      i\left(\frac{1}{2}\varGamma_a\sin(2\theta) -
        \dot{\theta}\right)\sin\varphi & &
      -i\varGamma_a\sin^2\theta & &
      i\left(\frac{1}{2}\varGamma_a\sin(2\theta) -
        \dot{\theta}\right)\cos\varphi 
      \\ \\
      -i\frac{1}{2}\varGamma_a\cos^2\theta\sin(2\varphi)-i\dot{\varphi}
      & &
      i\left(\frac{1}{2}\varGamma_a\sin(2\theta) + 
        \dot{\theta}\right)\cos\varphi & &
      \frac{1}{2}\Omega\tan\varphi  
      -i\varGamma_a\cos^2\theta\cos^2\varphi 
    \end{array}\right].
\end{equation}
\end{widetext}
Decay from the excited state $|e\rangle$ is neglected in order to
obtain more simple expressions. This approximation is justified by the
small involvement of the excited state and by assuming small decay
rate $\varGamma_e$. The state vector $\pmb{B}(t)$ is transformed by
the orthogonal rotation
\begin{equation}
  \pmb{O}(t)\!=\!\left[\begin{array}{ccc}
      \sin\varphi\sin\theta & \cos\theta & \cos\varphi\sin\theta
      \\
      \cos\varphi & 0 & -\sin\varphi  \\
      \sin\varphi\cos\theta & -\sin\theta & \cos\varphi\cos\theta
    \end{array}\right]
\end{equation}
according to
\begin{equation}
  \pmb{A}(t)\!=\!\pmb{O}^{-1}(t)\pmb{B}(t).
\end{equation}

In Eq.~(\ref{hamadi}) one can see that the decay from the target state
$|a\rangle$ affects the time evolution both of the dark state and the
bright states. More precisely, terms proportional to $\varGamma_a$
appear not only on the diagonal but also on the off-diagonal part of
the Hamiltonian. On one hand, these terms lead to a decrease of the
probability amplitudes for all of the three adiabatic states; on the
other hand, they contribute to nonadiabatic couplings which mix the
adiabatic states among themselves. If $\Omega_{{\rm
    max}}\!\gg\!\varGamma_a$ and the process is adiabatically slow,
then the diagonal elements dominate over the nondiagonal ones in
Eq.~(\ref{hamadi}). Then the states $|\psi_+(t)\rangle$ and
$|\psi_-(t)\rangle$ can be adiabatically eliminated by setting
$\dot{A}_+(t)\!=\!\dot{A}_-(t)\!=\!0$ in Eq.~(\ref{sch2}) and solving
the resulting algebraic equations for ${A}_+(t)$ and ${A}_-(t)$. Thus,
by inserting these solutions in the differential equation for
${A}_0(t)$ we find the final population $P_a(\varGamma_a)$ on the
state $|a\rangle$ to be
\begin{eqnarray}
  &&P_a(\varGamma_a)= \\ &&P_a
  \exp\left[-2\varGamma_a\int_{-\infty}^{\infty}\displaylimits
    \frac{\Delta^2\dot{\theta}^2\cos^2\theta+
      (\Omega^2/4+\dot{\varphi}^2)^2\sin^2\theta}
    {\Delta^2\varGamma_a^2\cos^4\theta+
      (\Omega^2/4+\dot{\varphi}^2)^2}dt\right],\nonumber
\end{eqnarray}
since $P_a(\varGamma_a)\!=\!|A_0(\infty)|^2$. The symbol $P_a$ denotes
the initial population on the state $|C\rangle$ which is given by
Eq.~(\ref{Ps}).  In the absence of decay $\varGamma_a\!=\!0$, we
recover the result derived earlier. For a significantly large decay
rate $\varGamma_a$, the target state $|a\rangle$ is emptied by the end
of the population transfer process. If the decay is accompanied by
emission of photons, then the signal will be proportional to the
population $P_a$.

In general, the initial state of the system cannot be described by a
state vector. In this case, our considerations in the preceding
paragraphs cannot be applied straightforwardly. However, some results
are still valid: In the Master equation (\ref{mast}) the projection of
the density operator $\hat\varrho$ on the decoupled state $|D\rangle$
is a conserved quantity during the transfer process. This implies that
the maximum amount of population which can be transferred to the
target state $|a\rangle$ is given by $P_a$ in Eq.~(\ref{Ps}).
Therefore, we have performed a numerical simulation, whose main
purpose is to verify whether this portion is indeed transferred. The
simulation may also allow us to study the adiabaticity of the process:
the excited state must be involved only minimally in the population
transfer. As a result we have verified numerically, for a wide range
of $\varGamma_a$ and for several initial density operators
$\hat\varrho_i$, that the population $\langle
C|\hat\varrho_f|C\rangle$ is zero and the population on the excited
state $|e\rangle$ is very small throughout the whole time.  The
quantity $\langle C|\hat\varrho_f|C\rangle$ is obtained from the
simulation in the presence of decay from the state $|a\rangle$.  The
parameters of the simulation have been chosen as follows: The atomic
system is described by $(\Delta, \varGamma_e, \varGamma_a )\!=\!(0.3,
0.1, 0.0 - 3.0)$; the pulses have been Gaussian with parameters
$(\Omega_{{\rm max}}, T, \tau)\!=(6.0, 2.0, 3.2)$, where $T$ denotes the
half-width of the pulses, and $\tau$ is the pulse delay. Time and
frequency are measured in arbitrary units.

In this way we have shown that combining a STIRAP process with a
population measurement on the target state of the population transfer,
yields the expectation value of the operator
\begin{eqnarray}
  \hat V\!&=&\!\cos^2\alpha\,|m\rangle\langle m|+
  \sin^2\alpha\,|n\rangle\langle n|
  \nonumber \\
  &+&\sin\alpha\,\cos\alpha(e^{i\beta}|m\rangle\langle n|+
  e^{-i\beta}|n\rangle\langle m|).
\end{eqnarray}
Performing a series of measurements of this kind by varying the angles
$\alpha$ and $\beta$, the density matrix elements
$(\hat\varrho_i)_{mm}$, $(\hat\varrho_i)_{nn}$, and
$(\hat\varrho_i)_{mn}$ can be unambiguously obtained. In
Refs.~\cite{vit1,vit2} a measurement procedure has been proposed where
the population from the ground state ensemble is transferred to an
excited state. This scheme seems more sensitive to the decay from the
excited state because it gets significantly populated during the process.
Therefore, the incoherent decay back to the ground states may
interrupt the coherent evolution which is required in this measurement
procedure as well.

In conclusion we have worked out a measurement procedure based on
stimulated Raman adiabatic passage to obtain the the density matrix
elements of a material system. The scheme is highly reliable due to
the robustness of the STIRAP process. In the beginning of the
measurement, a part of the density operator is expressed using two
orthogonal states: the coupled and the decoupled states, which are
defined by the two pump pulses. In this way a $2\times2$ block of the
density matrix is addressed. The coupled part of the density operator
is then transferred to an auxiliary state. We have shown that by
measuring the population on the auxiliary state for some
configurations of the pump pulses, the $2\times2$ block of the density
matrix can be determined.

The measurement procedure presented here can be used efficiently when
it is not necessary to obtain the complete density operator, but only
few elements are needed. We propose the implementation of this
scheme in several microscopic quantum systems, e.g. electronic states
of atoms, vibronic states of molecules etc..

This work was supported by the European Union Research and Training
Network COCOMO, contract HPRN-CT-1999-00129.

\newpage

\end{document}